\documentclass[prb,aps,twocolumn]{revtex4}
\usepackage{amsmath,epsfig,color,verbatim}
\begin{document}
\title{Electronic spectrum in cuprates within the  $\bf p$-$\bf d$ Hubbard
model} %
\author{N.M.Plakida$^{a,b}$ and V.S. Oudovenko$^{a,c}$  }
\address{$^a$Joint Institute for Nuclear Research, 141980 Dubna,  Russia\\
$^b$Max-Planck-Institut f\"ur Physik komplexer Systeme, D-01187 Dresden,
Germany\\
$^c$Rutgers University, Piscataway, New Jersey 08854, USA}

\date{\today}

\begin{abstract}
A microscopic theory for electronic  spectrum of the CuO$_2$ plane
within an effective $p$-$d$ Hubbard model  is proposed. Dyson
equation for the one-electron Green function in terms of the
Hubbard operators is derived  which is  solved self-consistently
for the self-energy evaluated in the noncrossing approximation.
Electron scattering on spin fluctuations induced by kinematical
interaction is described by a model dynamical spin susceptibility.
Doping and temperature dependence of electron dispersions,
spectral functions, the Fermi surface and the coupling constant
$\lambda$ are studied.
\end{abstract}


\maketitle

%------------------------------------
Recent ARPES studies revealed a complicated character of electronic structure
in the copper oxide superconductors which are believed to be caused by strong
electron correlations in cuprates,  as was originally suggested by
Anderson~\cite{Anderson87}. Below we report the results of electronic spectrum
calculations for an effective Hubbard model reduced from the $p$-$d$ model for
the CuO$_2$ plane in cuprates. In these studies for the first time  we go beyond
the mean-field approximation~\cite{Plakida95} or perturbation
approach~\cite{Krivenko05}. We have solved   the Dyson equation
self-consistently for the thermodynamic Green functions (GFs) and the
self-energy derived in the noncrossing approximation (NCA), $\,$ as has been
done by us for the $t$-$J$ model~\cite{Plakida99}.

\section{Model and Dyson equation}
\label{model}
 We  consider an effective $p$-$d$ Hubbard model for
one-hole states with energy $\varepsilon_1=\varepsilon_d-\mu$ and two-hole
$p$-$d$ singlet states with energy $\varepsilon_2=2\varepsilon_1+ U_{eff} $
where $\mu$ is the chemical potential and an effective Coulomb energy  $
U_{eff} = \Delta_{pd} = \epsilon_p-\epsilon_d $ (for details
see~\cite{Plakida95}):
\begin{eqnarray}
&& H = \varepsilon_1\sum_{i,\sigma}X_{i}^{\sigma \sigma} +
\varepsilon_2\sum_{i}X_{i}^{22} +  \sum_{i\neq j,\sigma}\, t_{ij}\,
\{X_{i}^{\sigma 0}X_{j}^{0\sigma}
 \nonumber \\
& + &  X_{i}^{\sigma 0}X_{j}^{0\sigma}
 + X_{i}^{2 \sigma}X_{j}^{\sigma 2} +2\sigma
(X_{i}^{2\bar\sigma}X_{j}^{0 \sigma} + {\rm H.c.})\},
 \label{m1}
\end{eqnarray}
where $X_{i}^{nm} = |in\rangle\langle im|$ are the Hubbard operators (HOs) for
4 states $\,n,\, m=|0\rangle , \; |\sigma\rangle,\; |2\rangle =|\uparrow
\downarrow \rangle $, $\sigma = \pm 1/2 $, $\bar\sigma=-\sigma$.  The
dispersion of holes is determined by the hopping parameters: $\, t_{ij} = t\,
\delta_{j, i\pm a_{x/y}} + t' \,\delta_{j, i \pm a_x \pm a_y} \,$ $(\, a_{x/y}
= a$ -  lattice constants). We take $\Delta_{pd} = 8 t \simeq 3.2$~eV  and
$\,t' = - \, 0.3\,t < 0$.
\par
By applying the Mori-type projection technique for the GFs ${\sf
G}_{ij\sigma}(t-t') = \langle\langle \hat X_{i\sigma}(t)\! \mid \! \hat
X_{j\sigma}^{\dagger}(t') \rangle\rangle$ in terms of the two-component HOs
($\, \hat X_{i\sigma}^{\dagger} = \{X_{i}^{2\sigma},\, X_{i}^{\bar\sigma 0}\})
$  an exact Dyson equation  was derived as described in~\cite{Plakida95} with a
self-energy (SE) as a many-particle GF. By using NCA for the SE, a closed
system of equations was obtained for the the GFs and the SE:
\begin{equation}
\tilde{G}_{1 (2)}({\bf q},\omega) =
 (\omega - \tilde{\varepsilon}_{1 (2)} ({\bf q})-
 \tilde{\Sigma}({\bf q},\omega))^{-1} \,  ,
 \label{m2}
\end{equation}
where  $\tilde{\varepsilon}_{1 (2)} ({\bf q})$ are  spectra for two bands given
by the matrix $\, \tilde{\varepsilon}\sb{ij} = \langle\{[\hat
X_{i\sigma},H],\hat X_{j\sigma}^{\dagger}\}\rangle \times \langle\{\hat
X_{i\sigma},\hat X_{i\sigma}^{\dagger}\} \rangle^{-1}$. The SE  is the same for
the both Hubbard bands:
\begin{eqnarray}
&&\tilde{\Sigma}({\bf k},\omega)=
  \frac{1}{\pi^2 N}\sum\sb{\bf q} | t({\bf q})|^{2}
  \! \int_{-\infty}^{\infty}\!\!\int_{-\infty}^{\infty}\frac{d\nu d z
N(z,\nu)}{\omega- z - \nu}
  \nonumber \\
& \times  & {\rm Im}\chi_{sc}({\bf k -q},\nu)\;
  {\rm Im}\, \{ \tilde{G}_{1}({\bf q},z)  + \tilde{G}_{2}({\bf q},z)
     \} \, .
 \label{m3}
\end{eqnarray}
Here the interaction  $t({\bf q})$ is determined by the hopping parameter $
t_{ij}$, $ N(z, \nu)= (1/2)(\tanh(z/2T)+ \coth(\nu/2T))$ and $\chi_{sc}({\bf
q},\nu) = \frac{1}{4}\langle\langle N_{\bf q}| N_{- \bf q} \rangle \rangle
_{\nu} + \langle \langle {\bf S}_{\bf q}|{\bf S}_{- \bf q}\rangle \rangle
_{\nu} $ is the charge-spin susceptibility where $N_{\bf q}$ is the number  and
${\bf S}_{\bf q}$ is the spin  operators.

\section{Results and discussions}

The system of equations (\ref{m2}), (\ref{m3}) was solved numerically for
various hole concentrations $n = 1+\delta = 2\langle X_{i}^{\sigma \sigma}
+X_{i}^{22} \rangle$ by using the Matsubara frequency representation at
temperature $T \simeq 0.03t \simeq 140$~K. Neglecting charge fluctuations, the
spin susceptibility was described by the model: ${\rm Im}\, \chi_{s}({\bf
q},\nu) = {\chi_0}/[1+ \xi^2 (1+ \gamma({\bf q}))] \, \tanh (\nu/2T) /
[1+(\nu/\omega_{s})^2]$ where $\xi$ is an antiferromagnetic (AF) correlation
length (in units of $a$), $\omega_{s}\simeq J = 0.4t$ is  spin-fluctuation
energy, and $\gamma({\bf q}) = (1/2)(\cos q_x + \cos q_y)$. The constant
$\chi_{0}= [3(1-|\delta|)/2 \omega_{s}] \{ ({1}/{N}) \sum_{\bf q} [
1+\xi^2[1+\gamma({\bf q})]^{-1} \}^{-1}$ is defined from the equation $\langle
{\bf S}_{i}{\bf S}_{i}\rangle = (3/4)(1 - |\delta|)$.
\par
\begin{figure}
\includegraphics[scale=.3]{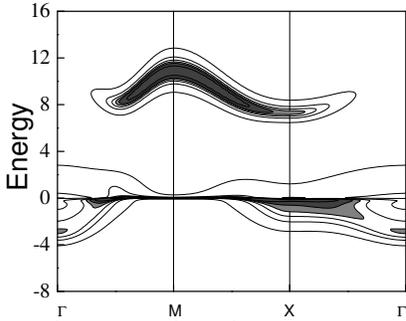}\\[-0.7cm]
 \caption{Dispersion curves  in units of $t$ along the symmetry
directions $\Gamma(0, 0)\rightarrow M(\pi,\pi) \rightarrow X (\pi,
0) \rightarrow \Gamma(0, 0)$ for $\delta = 0.1$ and temperature
$T= 0.03t$.}
 \label{figDA1-1a}
\end{figure}
\begin{figure}
\includegraphics[scale=.4]{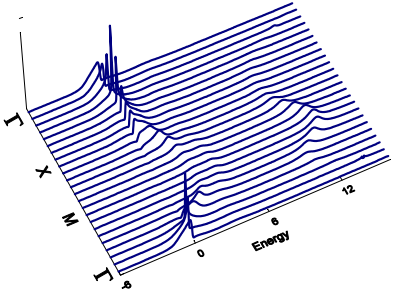}
\caption{Spectral function  in units of $t$ along the symmetry directions
$\Gamma(0, 0)\rightarrow M(\pi,\pi) \rightarrow X (\pi, 0) \rightarrow
\Gamma(0, 0)$ for $\delta = 0.1$ and temperature $T= 0.03t$.}
 \label{figDA1-1b}
\end{figure}

The dispersion curves given by  maxima of  spectral functions
$A({\bf k},\omega)= B_1({\bf k})\,\tilde{A}_{1}({\bf k}, \omega) +
B_2({\bf k})\, \tilde{A}_{2}({\bf k}, \omega)$, where $B_{1,
2}({\bf k})$ are the weights of the bands and
$\tilde{A}_{1(2)}({\bf k},\omega) = -(1/\pi) {\rm Im}
\tilde{G}_{1(2)} ({\bf k},\omega) $, were calculated for hole
doping $\delta = 0.1 - 0.3$.  The dispersion curves and the
spectral function for $\delta = 0.1\, (\xi = 2.5)$  reveal a
rather flat hole-doped band at the Fermi energy (FE) ($\omega =0$)
(Fig.~\ref{figDA1-1a}, Fig.~\ref{figDA1-1b}).

\begin{figure}
\centering
\includegraphics[scale=.4]{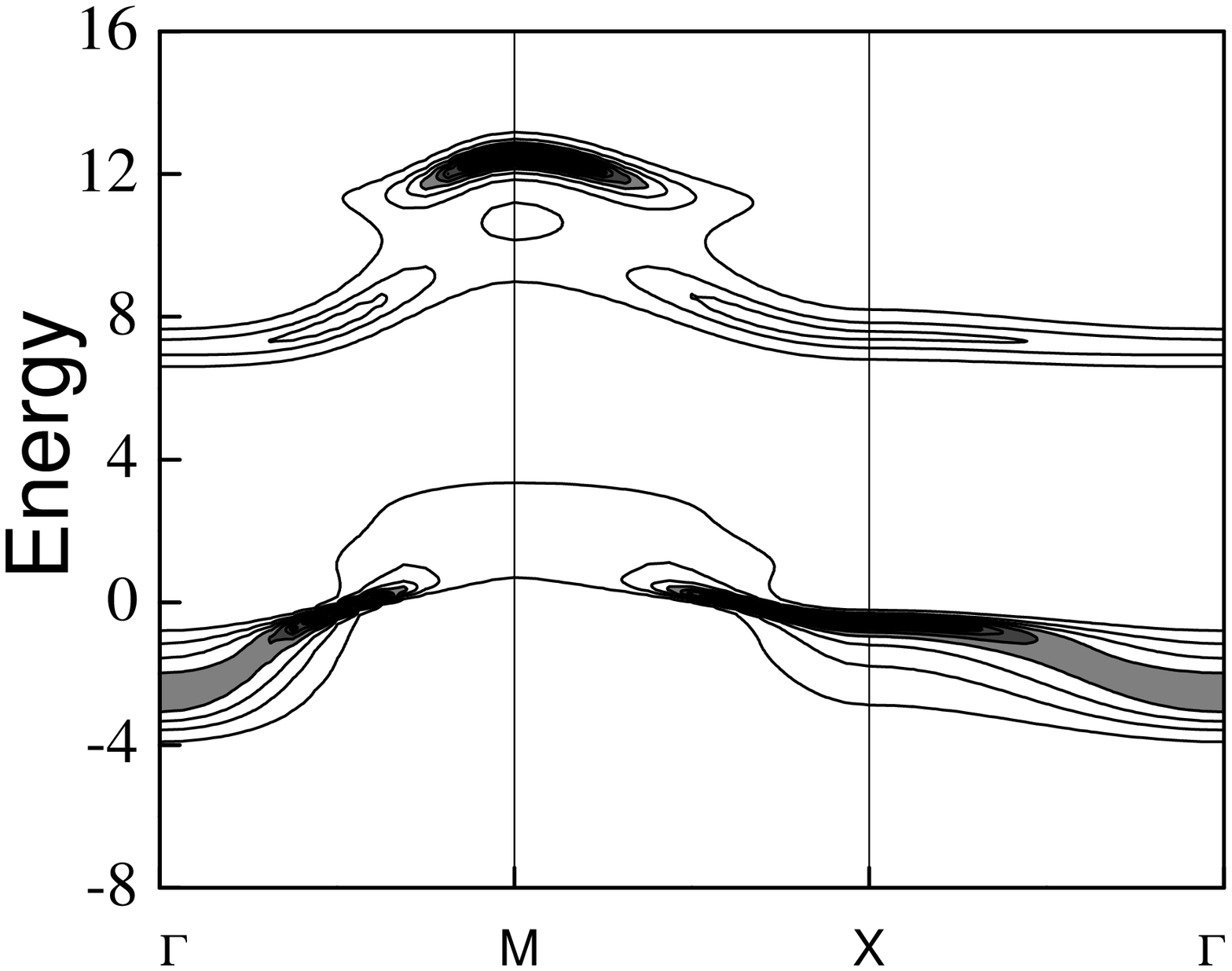}
 \caption{The same as Fig.~\ref{figDA1-1a} but for $T=0.3t$.}
 \label{figDA1-1Ta}
\end{figure}

\begin{figure}
\centering
\includegraphics[scale=.4]{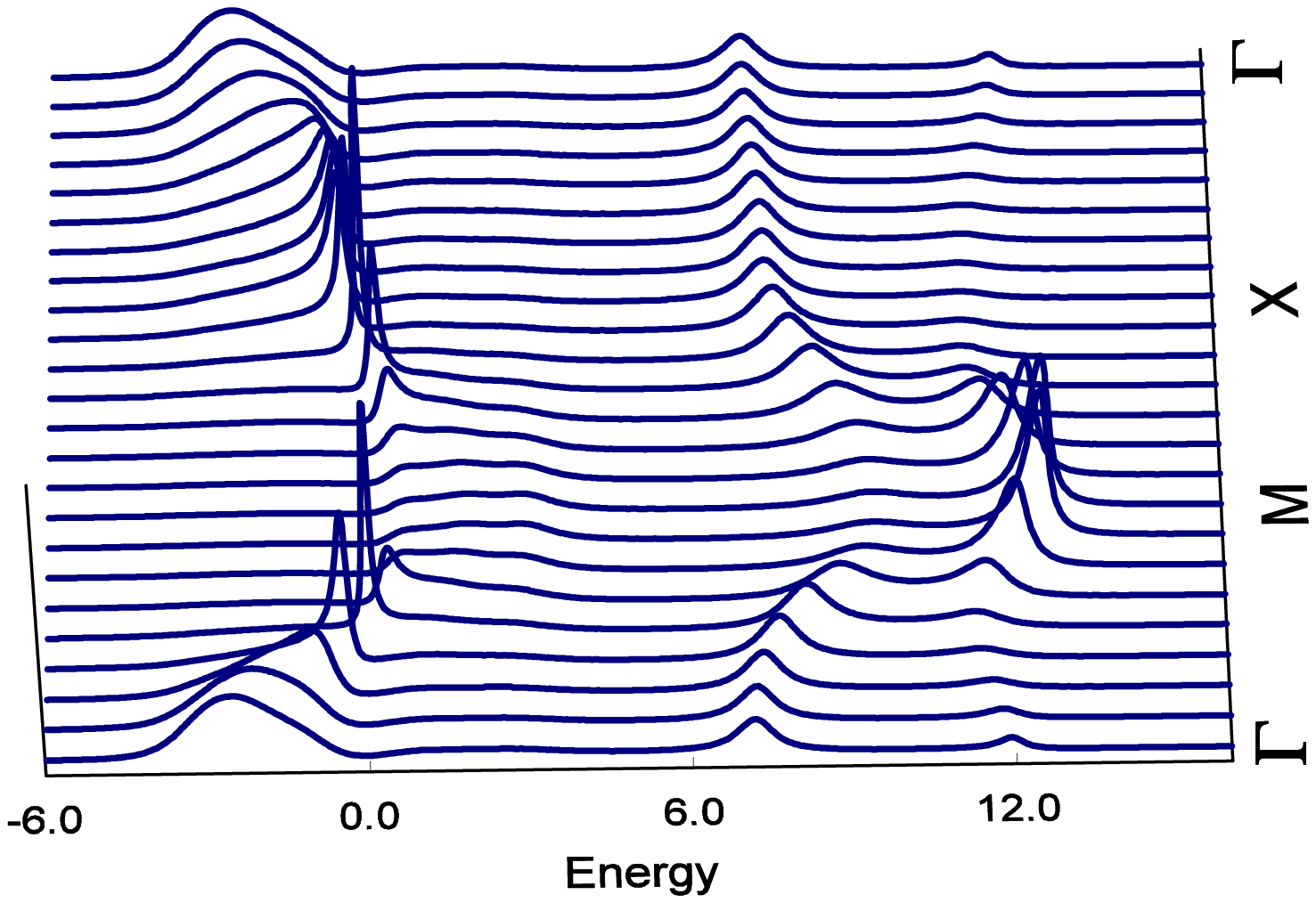}
 \caption{The same as Fig.~\ref{figDA1-1b} but for $T=0.3t$.}
 \label{figDA1-1Tb}
\end{figure}

At high temperature $T = 0.3t$ for $\delta = 0.1\, (\xi = 1.0)$
(Fig.~\ref{figDA1-1Ta}, Fig.~\ref{figDA1-1Tb}) or in the overdoped
region ($\delta = 0.3 $) the dispersion becomes much larger which
proves  a strong influence of AF spin-fluctuations on the
\begin{figure}[!t]
\includegraphics[scale=.4]{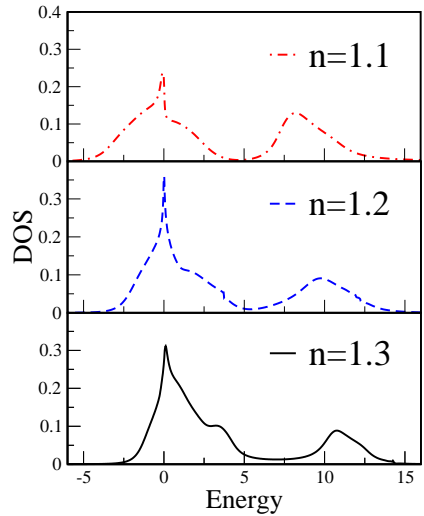}
 \caption{Electronic density of states.}
 \label{figDOS}
\end{figure}
\begin{figure}[!t]
\includegraphics[scale=.3]{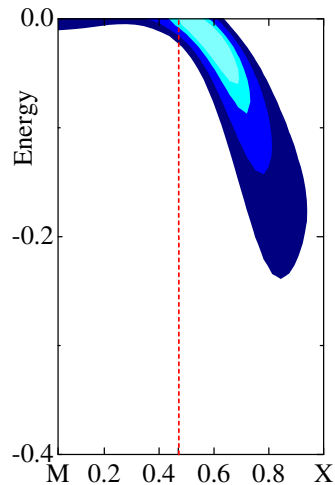}
 \caption{ $A({\bf k}, \omega)$ in the $M \rightarrow X $
  direction at the Fermi level crossing for $\delta = 0.1$.}
 \label{figSE}
\end{figure}
\begin{figure}[!htb]
\includegraphics[scale=.3]{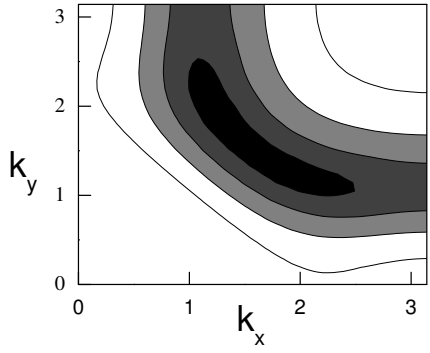}
 \caption{ $A({\bf k},\omega =0)$ on the FS for $\delta = 0.1$ .}
 \label{figF1}
\end{figure}
\begin{figure}[!htb]
\includegraphics[scale=.3]{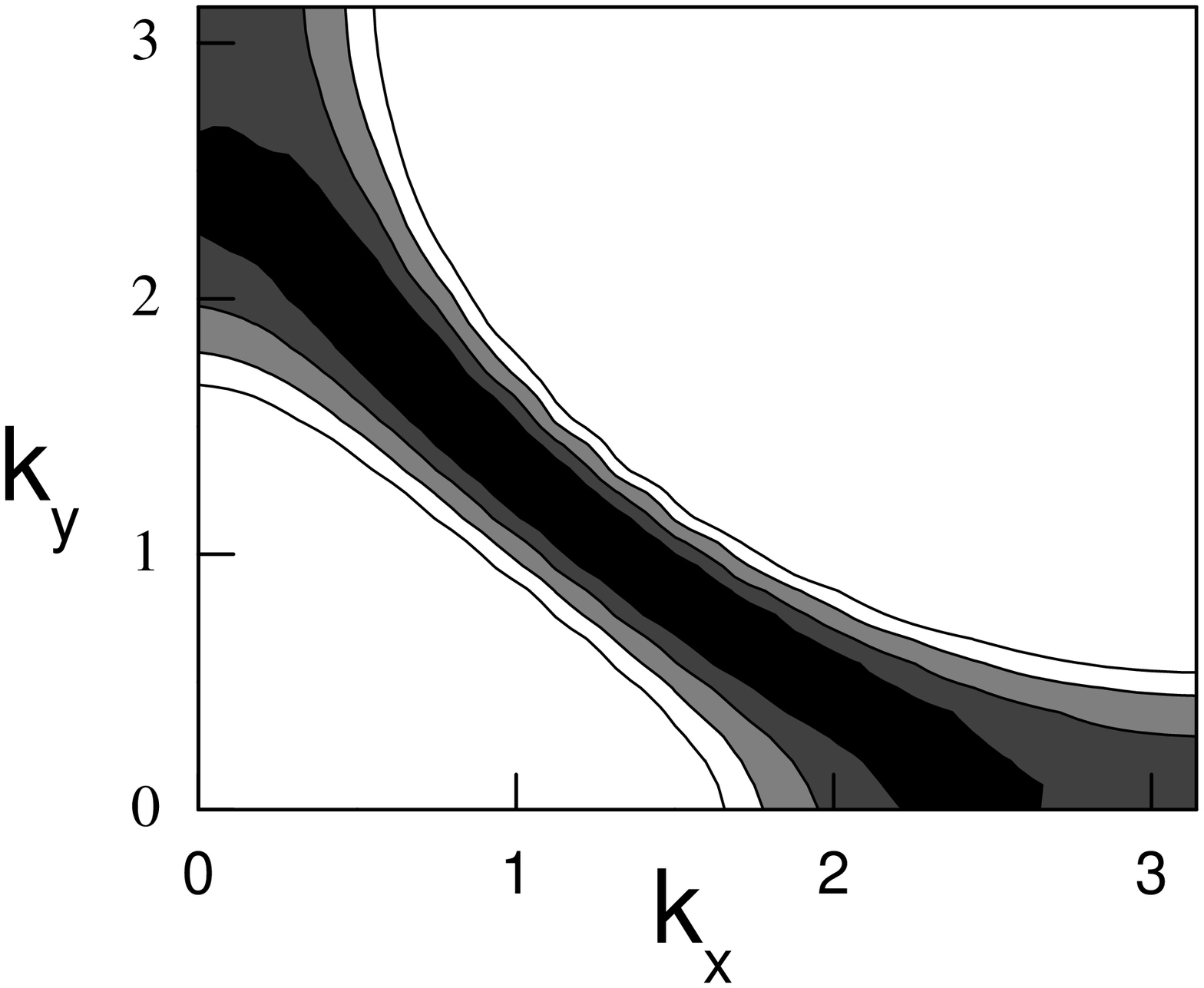}%
 \caption{The same as Fig.~\ref{figF1} but for $\delta = 0.2$.}
 \label{figF2}
\end{figure}
\begin{figure}[!htb]
\includegraphics[scale=.3]{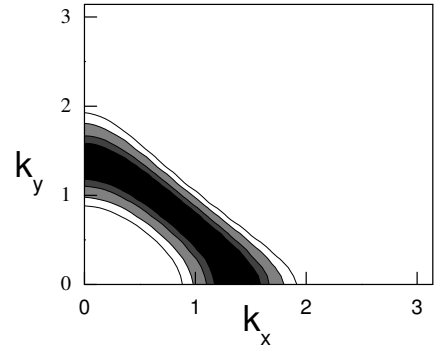}
 \caption{The same as Fig.~\ref{figF1} but for  $\delta = 0.3$.}
 \label{figF3}
\end{figure}

With doping, the density of states (DOS) shows a weight transfer
from the upper to the lower band as shown in Fig.~\ref{figDOS},
left panel.  The self-energy $\tilde{\Sigma}({\bf k}, \omega)$
reveals an  appreciable variation with ${\bf k}$ and doping close
to the Fermi level. Figure~\ref{figSE}  shows a change of
dispersion (kink) in the $M \rightarrow X $ direction at the Fermi
level crossing for $\delta = 0.1$. For the coupling constant we
get an estimation $\lambda = v_{\rm F}/ v_{0} -1 \simeq 2.4$
($\lambda  \simeq 0.7$ for $\delta = 0.3$). The FS changes  from a
hole arc-type at $\delta = 0.1$ to an electron-like one at $\delta
=0.3$ (Fig.~\ref{figF1}  -- Fig.~\ref{figF3}).%\\[-2cm]

To conclude,  the  microscopic  theory based on HO technique for
the effective $p$-$d$ model (\ref{m1}) provides an explanation for
doping and temperature dependence of electronic spectrum in
cuprates which is controlled by the AF spin correlations.
Superconducting pairing in the model beyond the weak coupling
approximation~\cite{Plakida03} will be considered elsewhere.
%\parbox[h]{16cm}{

\end{document}